\documentclass[letterpaper]{article}

\usepackage[T1]{fontenc}

\usepackage{geometry}
\usepackage{setspace}
\usepackage{amsmath}
\usepackage{siunitx}
\usepackage{amssymb}
\usepackage{graphicx}
\usepackage{hyperref}
\usepackage{float}
\usepackage{booktabs}   
\usepackage{multirow}
\usepackage{amsmath}
\usepackage{xcolor}
\usepackage{makecell}
\hypersetup{hypertex=true,
colorlinks=true,
linkcolor=blue,
anchorcolor=blue,
citecolor=blue}
\usepackage{ragged2e}
\justifying

\usepackage{achemso}

\usepackage{graphicx}
\usepackage{float}
\newfloat{scheme}{htbp}{los}
\floatname{scheme}{Scheme}
\floatname{chart}{Chart}
\newfloat{graph}{htbp}{loh}
\usepackage{chemformula} 
\usepackage[version = 4]{mhchem} 

\usepackage{authblk}
\author[1]{ShuaiYu Wang$^{\dagger}$}
\author[1]{Sheng Chen$^{\dagger}$}
\author[1,2]{Xiao-Ping Li}
\affil[1]{Research Center for Quantum Physics and Technologies, School of Physical Science and Technology, Inner Mongolia University, Hohhot, 010021, China}
\author[1,3]{Lei Wang*}
\affil[2]{Key Laboratory of Semiconductor Photovoltaic Technology and Energy Materials at Universities of Inner Mongolia Autonomous Region,
Inner Mongolia University, Hohhot 010021, China.}
\affil[3]{Inner Mongolia Key Laboratory of Microscale
Physics and Atom Innovation, Inner Mongolia University, Hohhot 010021, China.}

\title{Symmetry-Selective Strain Control of Spin--Momentum Locking and Spin Transport in Two-Dimensional Pentagonal Altermagnets}
\date{*Email: lwang@imu.edu.cn}

\begin{document}

\maketitle

\begin{abstract}
Altermagnets are compensated magnets featuring momentum-dependent nonrelativistic spin splitting generated by nontrivial operations connecting opposite-spin sublattices. A direct symmetry-based route to control this spin splitting is to modify the real-space operations that define the altermagnetic spin--momentum locking (SML). Here, we develop a strain-resolved symmetry framework for two-dimensional pentagonal altermagnets, classifying whether uniaxial and shear strain tensors preserve, reconstruct, or eliminate the SML. Using the above criterion combined with first-principles screening, we identify 94 stable altermagnetic candidates from 3330 materials. These candidates cover all type-III spin Laue groups of orthorhombic lattices and are classified into three strain-response types: Type-I preserves the SML; Type-II reconstructs the SML through partial symmetry breaking while retaining essential altermagnetic features; and Type-III destroys the altermagnetic SML. Representative materials further demonstrate this classification: ferroelastic $\alpha$-CoS$_2$ exhibits ferroelastically switchable SML and reverses the sign of the off-diagonal spin conductivity; shear-strained \(\alpha\)-CoP\(_2\) undergoes a \(g\)- to \(d\)-wave reconstruction of the SML, activating off-diagonal spin conductivity; and uniaxially strained FeSSe realizes strain-selected spin-valley transport. This work provides theoretical and material guidance for strain-controlled transport in two-dimensional orthorhombic altermagnets.
\end{abstract}

\section*{Keywords}

altermagnetism; strain engineering; pentagonal monolayers; spin transport; ferroelasticity; spin-valley coupling

\section{Introduction}
Altermagnets are collinear compensated magnets that exhibit momentum-dependent nonrelativistic spin splitting despite zero net magnetization \cite{song_altermagnets_2025,wei2024crystal}. This property distinguishes them from conventional spin-degenerate antiferromagnets while retaining the absence of stray magnetic fields characteristic of compensated magnetic order \cite{bai_altermagnetism_2024}. Unlike conventional antiferromagnetic spin degeneracy, altermagnetic spin splitting is governed by spin-group symmetry, which combines magnetic sublattices, spin-space operations, and real-space crystalline operations \cite{SSG1,SSG2}. When opposite-spin sublattices are connected by rotations or mirror operations, spin-up and spin-down bands can split in a momentum-dependent manner \cite{SSG3,AM-PRX-2022,AM-PRX2-2022}. Consequently, spin-dependent transport (spin current) in altermagnets is constrained by the symmetry operations that relate opposite-spin sublattices \cite{feng2022anomalous,bai2022observation,gonzalez2021efficient,shao2021spin,2022giant,SSE-SNE-magnon1,SSE-SNE-magnon2,SSE-magnon}. Therefore, a promising route to control altermagnetic transport is to gain control over the real-space symmetry operations that generate the altermagnetic spin-momentum locking (SML) \cite{2021multifunctional,2020crystal,FESAM-Liu,wang2025two,krempasky_altermagnetic_2024,badura2025observation,zhang_crystal-symmetry-paired_2025,he2024quasi,xu2025altermagnetic,leon2025hybrid,xie2026,qu2025altermagnetic,kong2019spin,Zhang2026}.

Existing routes for manipulating altermagnetic properties mainly focus on tuning the occupation, orientation, or interfacial environment of spin-split bands \cite{zhou_manipulation_2025,chen2025probing}. A more direct strategy is to modify the crystalline operations themselves. Strain provides such a route because it directly changes lattice constants, bond angles, and orbital hybridization and can selectively preserve, remove, or transform the real-space operations entering the spin-group symmetry \cite{bai_altermagnetism_2024,gao2025ai,fender2025altermagnetism,leon2025hybrid,waichang1,waichang2}. As a result, strain can alter not only the magnitude of the spin splitting but also the symmetry relations between opposite-spin sublattices \cite{fan2025valley,guo2024valley,leon2025hybrid}. Recent studies have shown that strain can tune spin splitting, spin--momentum locking, spin-valley polarization, and spin-transport responses in altermagnets \cite{fan2025valley,guo2024valley,li2025altermagnetism,huang2025spin,ding2025ferroelastically,ma2025bipolar,1chen,2chen,li2026manipulating}. However, most investigations have focused on a specific compound, a single deformation path, or an individual transport quantity. A unified strain-resolved symmetry criterion is still lacking for determining how different components of the strain tensor modify opposite-spin sublattice-connecting operations and, consequently, altermagnetic spin splitting and spin-resolved transport.

Two-dimensional pentagonal lattices provide a suitable platform for developing such a strain-resolved symmetry framework \cite{penta4}. Compared with high-symmetry honeycomb lattices, pentagonal networks usually possess lower in-plane symmetry, pronounced structural anisotropy, and multiple deformation channels \cite{penta3,penta2024}. As a result, different strain-tensor components, such as shear and uniaxial strain, are generally symmetry inequivalent and can act on different crystalline operations \cite{penta1,penta2}. In addition to continuous strain perturbations, several pentagonal monolayers host ferroelastic or quasi-ferroelastic structural variants, allowing finite strain to switch the system between degenerate or nearly degenerate domains \cite{guo2025mechanically,ding2025ferroelastically}. Such ferroelastic switching should be distinguished from symmetry-lowering strain: it may preserve the same point-group class and equivalent opposite-spin sublattice-connecting operations, while rotating the principal axes, spin-splitting pattern, and spin-conductivity tensor \cite{ding2025ferroelastically,huang2025spin,huang2026spin}. Therefore, pentagonal altermagnets offer a natural setting in which continuous strain tensors and finite ferroelastic domain switching can be treated within one symmetry framework to control SML and spin-current responses \cite{sodequist2024two,qu2025altermagnetic,jiang2026interplay,zhang2026lifshitz}.

In this work, we establish a strain-resolved symmetry framework for two-dimensional pentagonal altermagnets. We treat shear strain, uniaxial strain, and finite ferroelastic switching as symmetry-distinct perturbations and classify their effects according to whether the altermagnetic SML is preserved, reconstructed, or destroyed. Applying this criterion to ten parent space groups and first-principles-screened pentagonal monolayers yields 94 altermagnetic candidates, which cover all nontrivial spin Laue groups of orthorhombic lattices. We systematically present the evolution of all nontrivial spin Laue groups of orthorhombic lattices under different strain conditions. Three typical materials further demonstrate how this symmetry classification translates into functional responses: ferroelastic \(\alpha\)-CoS\(_2\) exhibits switching of altermagnetic SML, \(\alpha\)-CoP\(_2\) shows shear-driven reconstruction of spin-split bands and activation of off-diagonal spin conductivity, and FeSSe realizes uniaxial-strain-selected spin-valley transport. These results offer theoretical and material guidance for strain-tensor control of the altermagnetic SML and for mechanically tunable spin transport in two-dimensional altermagnetic orthorhombic lattices.

\section{Results and discussion}
We establish a strain-resolved symmetry framework for mechanically controlling altermagnetic spin splitting and spin transport in two-dimensional pentagonal altermagnets. In this framework, an in-plane strain tensor acts as a symmetry-selective perturbation to the crystalline operations connecting opposite-spin sublattices, thereby determining whether the spin–momentum locking relationship, hereafter referred to as SML, is preserved, reconstructed, or completely broken. By combining this symmetry criterion with high-throughput first-principles screening and spin-resolved transport calculations, we build a direct correspondence among strain-tensor symmetry, altermagnetic SML, and mechanically tunable spin conductivity.

\subsection{Symmetry Classification of Strain-Controlled Altermagnetism in Two Dimensions}

\begin{figure}[ht]
  \includegraphics[width=1\linewidth]{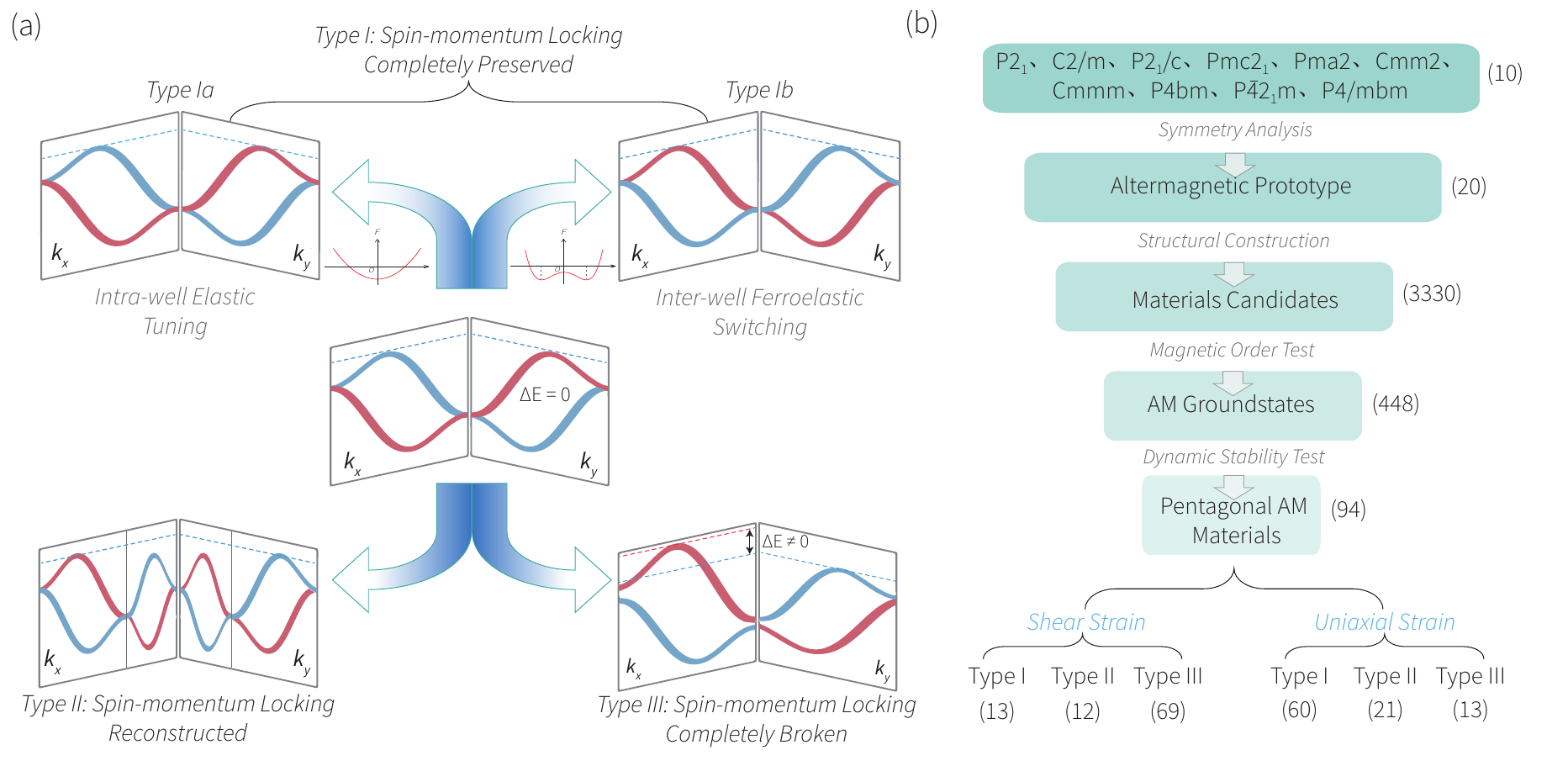}
  \caption{Strain-controlled SML and material screening in  two-dimensional pentagonal altermagnets. (a) Schematic of the classification of strain-induced SML evolutions under uniaxial and shear strain. Notably, Type I is further divided into two subcategories: Type Ia (intra-well elastic tuning) and Type Ib (inter-well ferroelastic switching). (b) Schematic of the high-throughput screening workflow for pentagonal altermagnetic materials. Also shown is the classification of strain-induced SML evolution under shear and uniaxial strain corresponding to Table \ref{tab:strain_final}, along with the number of materials associated with each classification.}
  \label{Fig1}
\end{figure}

We first establish a strain-resolved symmetry criterion to determine how in-plane strain modifies nonrelativistic altermagnetic spin splitting in two-dimensional compensated altermagnets. For collinear antiferromagnets, the nontrivial spin group takes the form $[E\parallel H] + [C_2\parallel AH]$. The operations in the subgroup $H$ connect sites with the same spin orientation, while those in the coset $AH$ connect sites with opposite spin orientations. In two-dimensional systems, additional spin-group operations, namely [$C_2 \parallel \mathcal{P}$], [$C_2 \parallel \tau$], [$C_2 \parallel m_z$], and [$C_2 \parallel C_{2z}$], can further protect spin-degenerate band structures. Therefore, for two-dimensional altermagnets, the allowed nontrivial spin Laue groups are restricted to a finite set, denoted here as $\mathcal{R}_{\mathrm{2D}}=\{{}^22/{}^2m_x,\; {}^2m{}^2m{}^1m,\; {}^24/{}^1m,\; {}^24/{}^1m{}^2m{}^1m,\; {}^1\bar{3}{}^2m,\; {}^14/{}^1m{}^2m{}^2m,\; {}^16/{}^1m{}^2m{}^2m \,\}$ \cite{spg}. For a given material, we denote the parent spin Laue group before strain as \(\mathcal{R}_0\), and the corresponding set of real-space operations connecting opposite-spin sublattices as \(\mathcal{O}_0^{\uparrow\downarrow}\). After applying an in-plane strain tensor \(\varepsilon\), the remaining spin Laue group is denoted as \(\mathcal{R}_\varepsilon\) and the surviving opposite-spin connecting operations is denoted as \(\mathcal{O}_\varepsilon^{\uparrow\downarrow}\), respectively. The operations that survive under strain are those that leave the strain tensor invariant:
\begin{equation}
\mathcal{O}_\varepsilon^{\uparrow\downarrow}
=
\left\{
g \varepsilon=\varepsilon
\mid
g\in \mathcal{O}_0^{\uparrow\downarrow}
\right\},
\label{eq:strain_invariant_operations}
\end{equation}
where $g$ is the real-space operation. Among these surviving operations, we further define \(\mathcal{O}_{\varepsilon,\mathrm{AM}}^{\uparrow\downarrow}\) as the subset of nontrivial opposite-spin sublattice-connecting operations that support an altermagnetic SML.

This symmetry criterion classifies the strain response into three types:
\begin{equation}
\begin{aligned}
\mathrm{Type\ I}:&\quad 
\mathcal{R}_\varepsilon=\mathcal{R}_0\in\mathcal{R}_{\mathrm{2D}},
\quad
\mathcal{O}_{\varepsilon,\mathrm{AM}}^{\uparrow\downarrow}
=
\mathcal{O}_{0,\mathrm{AM}}^{\uparrow\downarrow},\\
\mathrm{Type\ II}:&\quad 
\mathcal{R}_\varepsilon\neq\mathcal{R}_0,\quad
\mathcal{R}_\varepsilon\in\mathcal{R}_{\mathrm{2D}},
\quad
\mathcal{O}_{\varepsilon,\mathrm{AM}}^{\uparrow\downarrow}
\subset
\mathcal{O}_{0,\mathrm{AM}}^{\uparrow\downarrow},\\
\mathrm{Type\ III}:&\quad 
\mathcal{R}_\varepsilon\notin\mathcal{R}_{\mathrm{2D}},
\quad
\mathcal{O}_{\varepsilon,\mathrm{AM}}^{\uparrow\downarrow}
=
\varnothing .
\end{aligned}
\label{eq:strain_response_types}
\end{equation}
Type I corresponds to a symmetry-preserved response, in which strain retains the parent spin Laue group and the nontrivial opposite-spin sublattice-connecting operations responsible for altermagnetic spin splitting. Consequently, the altermagnetic SML is preserved. In ferroelastically coupled altermagnets, this response can induce switching between symmetry-related structural variants, resulting in a reversal of spin polarization without altering the underlying SML type.
Type II corresponds to a symmetry-reconstructed response. In this case, strain reduces the parent symmetry and removes some of the nontrivial operations that connect opposite-spin sublattices. Nevertheless, the strained structure still belongs to a two-dimensional altermagnetic spin Laue group. The altermagnetic SML is therefore reconstructed rather than destroyed. This reconstruction typically changes the momentum-space symmetry of the spin splitting, for example from an \(i\)-wave to a \(g\)-wave pattern, or from a \(g\)-wave to a \(d\)-wave pattern.
Type III corresponds to a symmetry-broken response. Here, the strained structure no longer belongs to the set of two-dimensional altermagnetic spin Laue groups and all symmetry operations connecting opposite-spin sublattices in the parent phase are completely broken. 

In this work, we focus on two elementary in-plane strain channels: uniaxial strain with transverse relaxation and pure shear strain. They are represented by
\begin{equation}
\varepsilon_{\mathrm{uni}}=
\begin{pmatrix}
\varepsilon_{xx} & 0\\
0 & \varepsilon_{yy}
\end{pmatrix},
\quad
\varepsilon_{\mathrm{shear}}=
\begin{pmatrix}
0 & \varepsilon_{xy}\\
\varepsilon_{yx} & 0
\end{pmatrix}.
\label{eq:strain_tensors}
\end{equation}
For the uniaxial strain channel, one in-plane lattice constant is externally varied, while the perpendicular lattice constant is allowed to relax through the Poisson response. The shear channel describes an off-diagonal deformation of the two-dimensional lattice. Because strain is a second-rank tensor, each strain channel is invariant only under a subgroup of the parent point group. Different strain tensors therefore selectively retain, reconstruct, or remove the crystalline operations that connect opposite-spin sublattices, providing a symmetry-based route to control altermagnetic SML and the associated spin-resolved transport response in two-dimensional pentagonal altermagnets.

\subsection{Symmetry-Guided Screening of Pentagonal Altermagnetic Monolayers}

We next apply the strain-dependent symmetry-evolution criterion to two-dimensional orthogonal pentagonal lattices. A parent crystallographic space group is considered compatible with pentagonal altermagnetism when magnetic atoms occupying symmetry-related Wyckoff positions carry opposite spins and are connected by at least one nontrivial real-space operation, such as a rotation or mirror operation, that maps one spin sublattice onto the opposite-spin sublattice \cite{2024minimal}. On the basis of this criterion, we identify ten parent space groups that are compatible with pentagonal altermagnetism: \(P2_1\), \(C2/m\), \(P2_1/c\), \(Pmc2_1\), \(Pma2\), \(Cmm2\), \(Cmmm\), \(P4bm\), \(P\bar{4}2_1m\), and \(P4/mbm\). The corresponding space groups, spin Laue groups, strain-evolution types, and nontrivial opposite-spin sublattice-connecting operations are summarized in Table S1 in the Supporting Information.

We then performed high-throughput first-principles screening within these symmetry-allowed space groups. Starting from the 20 pentagonal prototypes, including 5 binary and 15 ternary structural templates shown in Figure S1 in the Supporting Information, elemental substitution generated 3330 candidate monolayers. For each candidate, four collinear magnetic configurations, namely FM, AM, AFM1, and AFM2, were fully relaxed and compared in energy, as illustrated in Figure S4 in the Supporting Information. This workflow yielded 448 magnetic candidates. Subsequent phonon calculations further identified 94 dynamically stable monolayers. Their space groups, spin Laue groups, and strain-response classifications are summarized in Figure \ref{Fig1}b and Table \ref{tab:strain_final}. Detailed structural information, relative energies of different magnetic configurations, and strain-evolution types are provided in Tables S2 and S3 in the Supporting Information. In particular, for different structures belonging to the same space group, we denote them as $\alpha$, $\beta$, and $\gamma$, respectively. The phonon spectra and electronic band structures of all dynamically stable monolayers are shown in Figures S2 and S3.

The resulting materials statistics reveal a pronounced tensor-selective strain response. One class of spin Laue groups preserves altermagnetic symmetry under uniaxial strain but loses it under shear strain, whereas another class exhibits the opposite behavior. For the \({}^{2}m_x\) spin Laue group, uniaxial strain preserves the relevant opposite-spin sublattice-connecting symmetry; therefore, the sublattice pairing and altermagnetic SML remain unchanged, corresponding to a Type I response. In contrast, shear strain removes the nontrivial symmetry operation required for altermagnetic SML. The strained system thus becomes a compensated collinear magnetic state that no longer satisfies the symmetry criterion for altermagnetism, corresponding to a Type III response.
The \({}^{2}m{}^{2}m{}^{1}m\) spin Laue group, in which opposite-spin sublattices are connected by two mutually perpendicular mirror operations, shows the opposite behavior. Shear strain preserves the relevant mirror-protected sublattice connection and maintains the altermagnetic SML, corresponding to Type I. By contrast, uniaxial strain breaks the required symmetry constraint and destroys the SML, leading to a Type III response.
For spin Laue groups such as \({}^{2}4/{}^{1}m\), where the opposite-spin sublattices are connected by a fourfold rotational operation, uniaxial strain lowers the rotational symmetry. If the strained structure retains or develops mirror-type operations that still connect opposite-spin sublattices, the system remains altermagnetic but with a reconstructed SML. This corresponds to a Type II response. Under shear strain, however, the required nontrivial sublattice-connecting operations are removed, leading to a Type III response.
For composite spin Laue groups combining mirror operations and higher-order rotational axes, such as \({}^{1}4/{}^{1}m{}^{2}m{}^{2}m\), both uniaxial and shear strains partially remove the original nontrivial opposite-spin sublattice-connecting operations. Nevertheless, sufficient symmetry remains to support altermagnetic order. These cases therefore exhibit a Type II response, characterized by a reconstruction of the original altermagnetic SML rather than its complete disappearance.

Overall, these results demonstrate that continuous lattice distortion provides a tensor-selective route to control pentagonal altermagnetism. By selectively preserving, reconstructing, or removing the crystalline operations that connect opposite-spin sublattices, uniaxial and shear strains determine the reduction of spin Laue symmetry, the evolution of sublattice pairing, and the resulting altermagnetic SML. The uneven distribution of stable compounds among different spin Laue groups further reflects chemical constraints, including the size of the substitution space, magnetic-energy competition, and dynamical stability.

\begin{table}[ht]
\centering
\caption{Summarized are 10 space groups (SG) capable of hosting altermagnetism, including the spin Laue group (SLG) for each space group's materials, the classification of spin–momentum locking evolution under strain, and the material counts. Specifically, space group No. 100 contains two altermagnetic configurations, corresponding to two distinct SLG.}
\setlength{\tabcolsep}{3pt}
\renewcommand{\arraystretch}{0.6}
\resizebox{\textwidth}{!}{%
\begin{tabular}{@{}l|l|cccc|ccc|c|ccc@{}}
\toprule
\multicolumn{2}{c|}{SLG} & \multicolumn{4}{c|}{${}^22/{}^2m_x$} & \multicolumn{3}{c|}{${}^2m{}^2m{}^1m$} & \multicolumn{1}{c|}{${}^24/{}^1m$} & \multicolumn{3}{c}{${}^14/{}^1m{}^2m{}^2m$} \\
\midrule
\multicolumn{2}{c|}{SG} & P2$_1$ & P2$_1$/c  & Pmc2$_1$ & Pma2 & P4bm-$\alpha$ & C2/m & Cmm2 & Cmmm & P4bm-$\beta$  & $P\bar{4}2_1m$ & P4/mbm \\
\midrule
\multirow{2}{*}{Shear}
& Type & III & III & III & III & III & I & I & I& II & II & II \\
& Numbers & 47 & 5 & 8 & 0 & 9 & 3 & 6 & 4&1 & 9 & 2 \\
\midrule
\multirow{2}{*}{Uniaxial}
& Type & I & I & I & I & II & III & III & III & II & II & II \\
& Numbers & 47 & 5 & 8 & 0 & 9 & 3 & 6 & 4&1 & 9 & 2 \\
\bottomrule
\end{tabular}%
}
\label{tab:strain_final}
\end{table}

\subsection{Representative Material Realizations of Strain-Controlled Altermagnetic Functionalities}

We now use three stable pentagonal monolayers---\(\alpha\)-CoS\(_2\), \(\alpha\)-CoP\(_2\), and FeSSe---to illustrate how the strain-resolved generator classification determines spin-split band structures and spin-current responses. \(\alpha\)-CoS\(_2\) belongs to space group \(P2_1/c\), representing a Type-I finite-strain ferroelastic response; \(\alpha\)-CoP\(_2\) belongs to space group \(P4/mbm\), exhibiting a Type-II response; and FeSSe belongs to space group \(Cmm2\), serving as a typical example of a Type-III uniaxial-strain response.

\subsubsection{Ferroelastic Switching of Altermagnetic SML in \(\alpha\)-CoS\(_2\)}

\begin{figure}[!htb]
\includegraphics[width=1\linewidth]{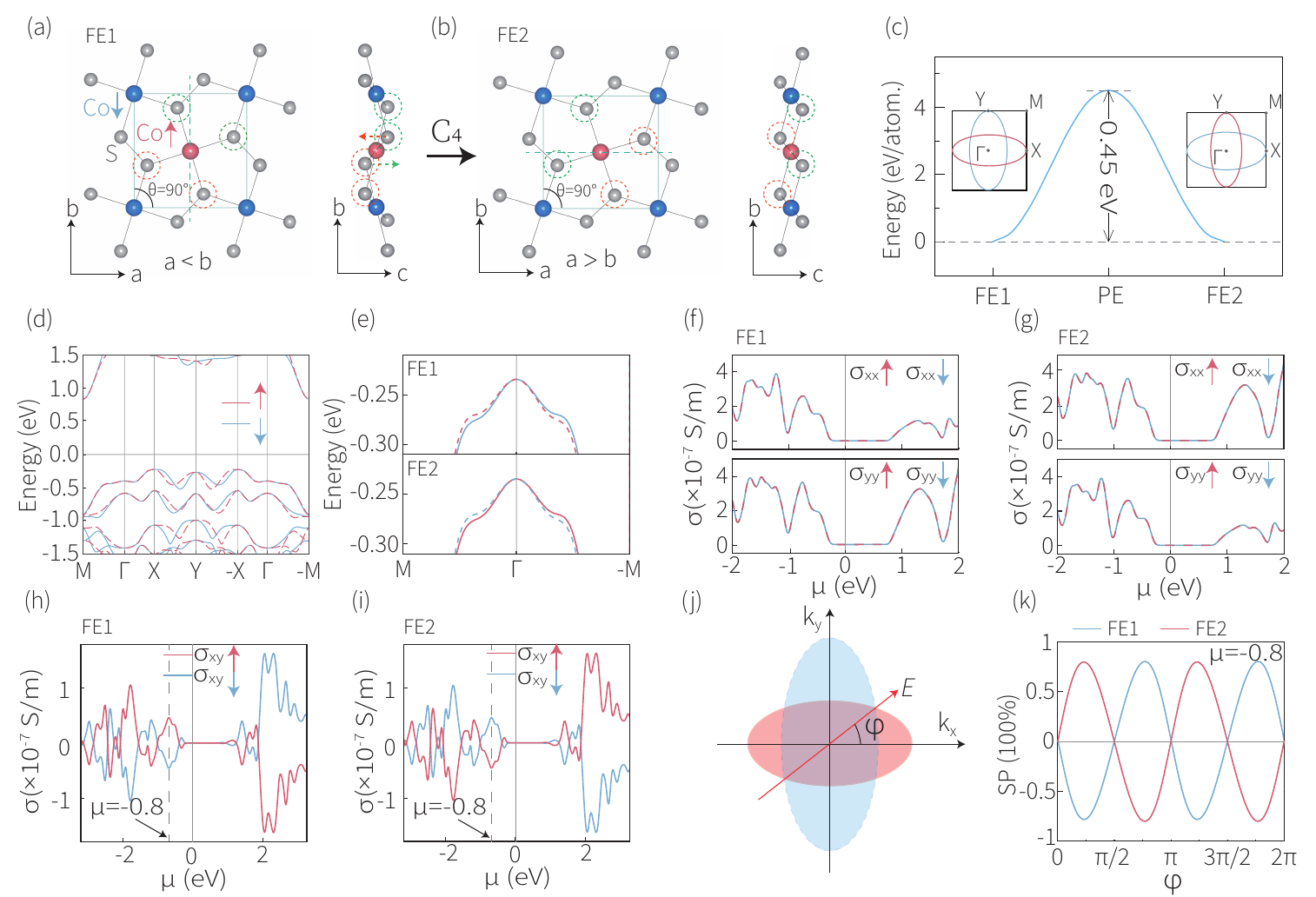}
\caption{Schematic illustration of ferroelastic altermagnetic pentagonal $\alpha$-CoS$_2$ monolayer and associated transport properties. (a) and (b) show the top and side views of the pentagonal $\alpha$-CoS$_{2}$ monolayer in the FE1 and FE2 states. (c) Energy barrier for ferroelastic switching estimated by the CI-NEB method. The following diagrams illustrate the SML of the two ferroelastic states. (d) and (e) The band structure and the top of valence band for the FE1 and FE2 states of $\alpha$-CoS$_{2}$. (f) and (g) Transverse ($\sigma^{s}_{xx}$) and longitudinal ($\sigma^{s}_{yy}$) spin-resolved conductivity as a function of chemical potential at 300 K for the FE1 and FE2 states, respectively. (h) and (i)  Off-diagonal ($\sigma^{s}_{xy}$) spin-resolved conductivity as a function of chemical potential at 300 K for the FE1 and FE2 states, respectively. (j) Schematic diagram illustrating the direction of the applied in-plane electric field. Red and blue represent the Fermi surface contours for spin-up and spin-down, respectively, and $\varphi$ is the angle between $k_x$ and the direction of the electric field. (k) Off-diagonal ($\sigma^{s}_{xy}$) spin polarization as a function of $\varphi$ with the Fermi energy fixed at -0.8 eV.}
  \label{Fig2}
\end{figure}

We first examine \(\alpha\)-CoS\(_2\), a pentagonal altermagnet belonging to space group \(P2_1/c\) and spin Laue group \({}^22/{}^2m_x\), which serves as a Type-Ib finite-strain ferroelastic case. As shown in Figure~\ref{Fig2}a,b, the monolayer has two degenerate ferroelastic variants, FE1 and FE2 states, characterized by \(a<b\) and \(a>b\), respectively. They are related by a ferroelastic domain operation equivalent to a \(90^\circ\) rotation of the in-plane axes. The bistability originates from the change in the local CoS\(_4\) coordination unit, in which two S atoms are displaced above and the other two below the Co plane.

The switching path between FE1 and FE2 states are shown in Figure~\ref{Fig2}c. The two variants are connected through a centrosymmetric paraelastic configuration with an energy barrier of approximately \(0.45\)~eV per unit cell, indicating a finite but accessible ferroelastic switching pathway. According to Aizu's symmetry theory of ferroelastic materials, different orientation variants are related by \(\mathcal{D}_{\text{FE}}\) operators, which belong to the paraelastic point group but not to the point groups of the orientation states \cite{aizu1,huang2026spin}. For \(\alpha\)-CoS\(_2\), the two ferroelastic variants FE1 and FE2 states are related by a domain operation \(\mathcal{D}_{\text{FE}} = C_4T\), which combines a \(90^\circ\) rotation and time-reversal symmetry. Since this domain operation is equivalent to interchanging the crystallographic \(x\) and \(y\) axes, it also maps \(k_x\) onto \(k_y\) in reciprocal space.

The key symmetry feature of ferroelastic switching in this altermagnet is that it does not remove the altermagnetic operations. The nontrivial operations connecting opposite-spin sublattices before and after switching are both
$\mathcal{O}^{\uparrow\downarrow}_{0}=\mathcal{O}^{\uparrow\downarrow}_{\varepsilon}=[C_{2x}, M_x].$
Thus, the opposite-spin-sublattice-connecting operation sets in the two variants are related by conjugation:
$\mathcal{O}_{\varepsilon}^{\uparrow\downarrow} = \mathcal{D}_{\mathrm{FE}} \mathcal{O}_{0}^{\uparrow\downarrow} \mathcal{D}_{\mathrm{FE}}^{-1}$. Therefore, \(\alpha\)-CoS\(_2\) remains in the same Type-I symmetry class across the ferroelastic transition, while the altermagnetic SML is rotated.
Figures~\ref{Fig2}d,e show the spin-resolved band structure of FE1, where altermagnetic spin splitting is observed along the high-symmetry paths $\Gamma \rightarrow $M and $\Gamma \rightarrow $-M. Along the \(k_x\) and \(k_y\) directions, the spin-up and spin-down bands become degenerate. This degeneracy arises from the combination of the pure spin operation \(C_2\) with the spatial operations \(C_{2x}\) and \(M_x\). For FE2, the same spin-splitting pattern is retained but rotated in momentum space, consistent with
$\Delta E_{\text{FE2}}(\mathbf{k}) = \Delta E_{\text{FE1}}(C_4T\,\mathbf{k}),$
where \(\Delta E(\mathbf{k})\) denotes the altermagnetic spin splitting. This domain rotation imposes a corresponding transformation on the spin-channel conductivity tensor.
As shown in Figures~\ref{Fig2}f,g, the longitudinal components are strongly anisotropic and are interchanged by ferroelastic switching, satisfying
$\sigma_{xx}^{\text{FE1}} = \sigma_{yy}^{\text{FE2}}, \sigma_{yy}^{\text{FE1}} = \sigma_{xx}^{\text{FE2}}.$
The equal longitudinal conductivities of the two spin channels reflect the compensated character of the magnetic order, whereas the directional anisotropy originates from the ferroelastic lattice distortion. The off-diagonal component provides a clearer spin-transport signature. Figures~\ref{Fig2}h,i show that the two spin channels satisfy:
$\sigma_{xy}^{\uparrow} = -\sigma_{xy}^{\downarrow},$
so the charge-like off-diagonal response cancels, while a pure spin-type transverse response remains. Upon switching from FE1 to FE2 states, the sign of this off-diagonal spin response is reversed, consistent with the $C_{4z}$ rotational symmetry of the crystal, which relates the two FE configurations.

Moreover, we calculated the anisotropic spin polarization of the two ferroelastic states at a chemical potential of \(-0.8\)~eV as a function of the in-plane field angle \(\varphi\). Figure~\ref{Fig2}j presents a schematic diagram of the direction of the external in-plane electric field. The result is shown in Figure~\ref{Fig2}k, the maximum polarization reaches \(80\%\). Moreover, the FE1 and FE2 variants give opposite spin-polarization patterns, demonstrating that ferroelastic domain switching reverses altermagnetic spin transport without generating net magnetization. In addition, we further predicted five other ferroelastically coupled altermagnetic materials, the results including the spin-resolved band structures and switching barriers are presented in Figure~S5 in the Supporting Information.

\subsubsection{Shear-Unlocked Transverse Spin Response in \(\alpha\)-CoP\(_2\)}

\begin{figure}[!ht]
\center
\includegraphics[width=0.95\linewidth]{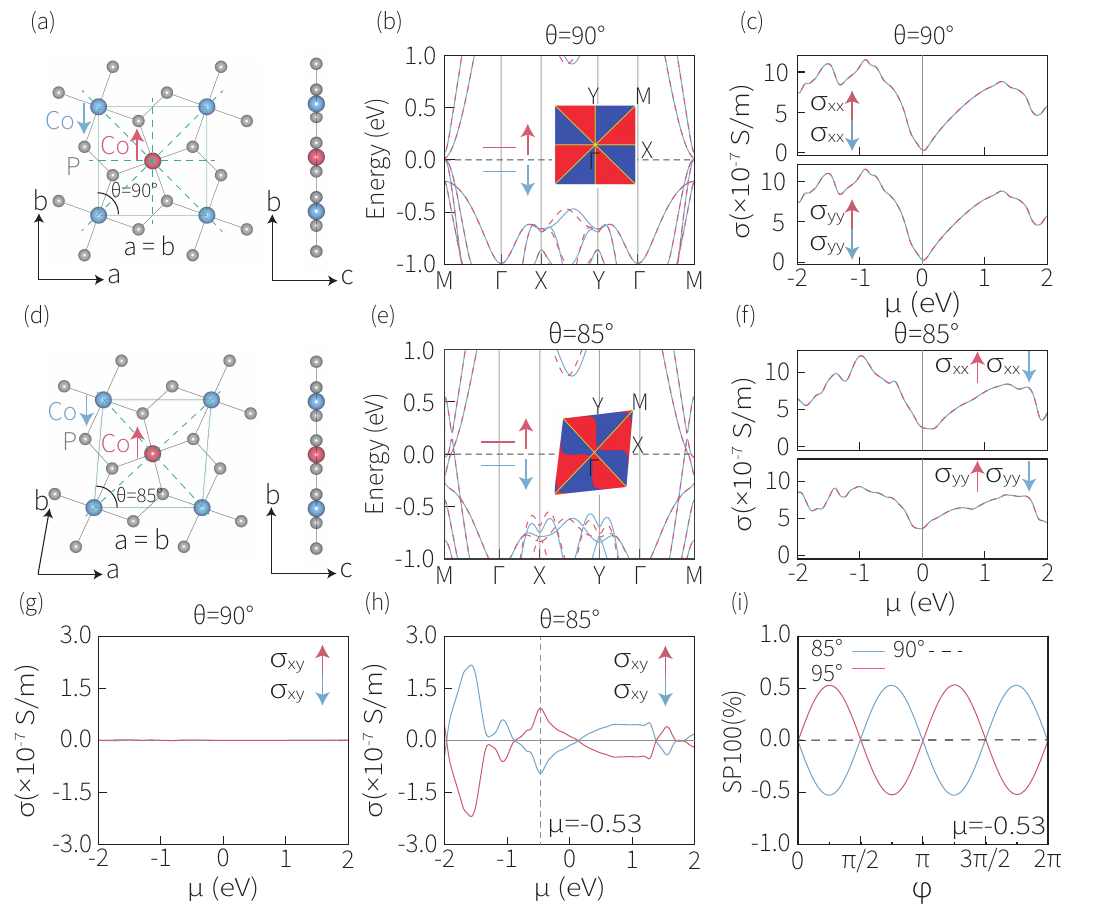}
\caption{Schematic illustration of strain-tunable electronic and spin transport properties in the $\alpha$-CoP$_2$ monolayer. (a) The crystal structure of $\alpha$-CoP$_{2}$ without strain. (b) and (c) The electronic band structure of $\alpha$-CoP$_{2}$ without strain and corresponding transverse ($\sigma^{s}_{xx}$) and longitudinal ($\sigma^{s}_{yy}$) spin-resolved conductivity as a function of chemical potential at 300 K of $\alpha$-CoP$_{2}$ monolayer, the inset in (b) shows a schematic of the $g$-wave altermagnetic Fermi surface. (d) The crystal structure of $\alpha$-CoP$_{2}$ after applying shear strain, where the angle between the  $a$ and $b$ becomes 85$^{\circ}$. (e) and (f) The electronic band structure and corresponding transverse ($\sigma^{s}_{xx}$) and longitudinal ($\sigma^{s}_{yy}$) spin-resolved conductivity as a function of chemical potential at 300 K of $\alpha$-CoP$_{2}$ monolayer with shear strain, the inset in (e) shows a schematic of the $d$-wave altermagnetic Fermi surface. (g) and (h) Off-diagonal ($\sigma^{s}_{xy}$) spin-resolved conductivity as a function of chemical potential at 300 K for the $\alpha$-CoP$_{2}$ without and with strain. (i) Off-diagonal ($\sigma^{s}_{xy}$) spin polarization as a function of $\phi$ with the Fermi energy fixed at -0.53 eV for the $\alpha$-CoP$_{2}$ monolayer with and without strain.}
\label{Fig3}
\end{figure}

\(\alpha\)-CoP\(_2\) is a Type-II shear-response case. In the unstrained phase, the lattice satisfies \(a=b\) and \(\theta=90^\circ\) (Figure~\ref{Fig3}a). This material crystallizes in the tetragonal space group \(P4/mbm\) (No.~127), with spin Laue group of \({}^14/{}^1m{}^2m{}^2m\). The \(C_{4z}\) symmetry connects four mutually perpendicular mirror planes and twofold rotation axes that exchange opposite-spin sublattices. Therefore, the full set of nontrivial symmetry operations connecting opposite-spin sublattices is: $\mathcal{O}_{0}^{\uparrow\downarrow} = [M_x, M_y, M_{xy}, M_{\overline{xy}}, C_{2x}, C_{2y}, C_{2xy}, C_{2\overline{xy}}].$
This symmetry analysis is corroborated by the band structure in Figure~\ref{Fig3}b. The bands show nonrelativistic, momentum-dependent altermagnetic spin splitting subject to the fourfold symmetry constraint. Electronic states with opposite spins remain degenerate along the high-symmetry lines:$\Gamma$ $\rightarrow$ X, $\Gamma$ $\rightarrow$ Y, $\Gamma$ $\rightarrow$ M, and $\Gamma$ $\rightarrow$ -M, while clear altermagnetic spin splitting is resolved along the high-symmetry path X $\rightarrow$ Y. The same symmetry also determines the form of the in-plane conductivity tensor. For each spin channel:
$\sigma_{xx}^{s} = \sigma_{yy}^{s}, \sigma_{xy}^{s} = 0,$
as confirmed in Figures~\ref{Fig3}c,g. Thus, the unstrained tetragonal phase allows longitudinal spin-resolved transport but forbids an off-diagonal spin-current response.

A pure shear deformation changes the lattice angle from \(\theta=90^\circ\) to \(\theta=85^\circ\), as shown in Figure~\ref{Fig3}d, corresponding to the shear strain tensor in Eq \ref{eq:strain_tensors}. The off-diagonal strain breaks the in-plane fourfold rotational symmetry, retaining only two mutually perpendicular mirror planes that exchange opposite-spin sublattices. Consequently, the spin Laue group transforms to \({}^2m{}^2m{}^1m\), and the nontrivial operations connecting opposite-spin sublattices become:
$\mathcal{O}_{\varepsilon}^{\uparrow\downarrow} = [M_{xy}, M_{\overline{xy}}, C_{2xy}, C_{2\overline{xy}}].$
Although nontrivial opposite-spin sublattice-connecting operations still exist, the reduced symmetry decreases the number of such generators and reconstructs the spin-splitting pattern, as shown in Figure~\ref{Fig3}e. Meanwhile, the spin--momentum locking is reduced from a \(g\)-wave-like to a \(d\)-wave-like form, as shown in the insets of Figures~\ref{Fig3}b,e. This places \(\alpha\)-CoP\(_2\) in the Type-II class: shear strain does not destroy altermagnetism but changes the form of the SML.

The symmetry reduction uniformly manifests as two correlated transport signatures: the anisotropy of the longitudinal conductivity ($\sigma_{xx}^{s} \neq \sigma_{yy}^{s}$, Figure~\ref{Fig3}f), which in turn gives rise to the spin-polarized current satisfying $\sigma_{xy}^{\uparrow} = -\sigma_{xy}^{\downarrow}$ (Figure~\ref{Fig3}h).
The sign of the response follows the sign of the shear distortion. Because \(\varepsilon_{xy}\) changes sign under the mirror operation that exchanges the two shear variants, the spin response is odd in shear: $\sigma_{xy}^{s}(+\varepsilon_{xy}) = -\sigma_{xy}^{s}(-\varepsilon_{xy}).$
Consistently, the \(\theta=85^\circ\) and \(\theta=95^\circ\) structures show opposite angular spin-polarization patterns in Figure~\ref{Fig3}i. Near the chemical potential of 0.53~eV, the spin polarization reaches about \(50\%\). Therefore, shear strain lifts the crystalline symmetry constraints that enforce $\sigma_{xx}^{s} = \sigma_{yy}^{s}$ in the tetragonal phase, thus unlocking a spin-polarized current channel in $\alpha$-CoP$_2$.

\subsubsection{Uniaxial-Strain Selection of Spin-Valley Conducting Channels in FeSSe}

\begin{figure}[ht]
  \includegraphics[width=0.95\linewidth]{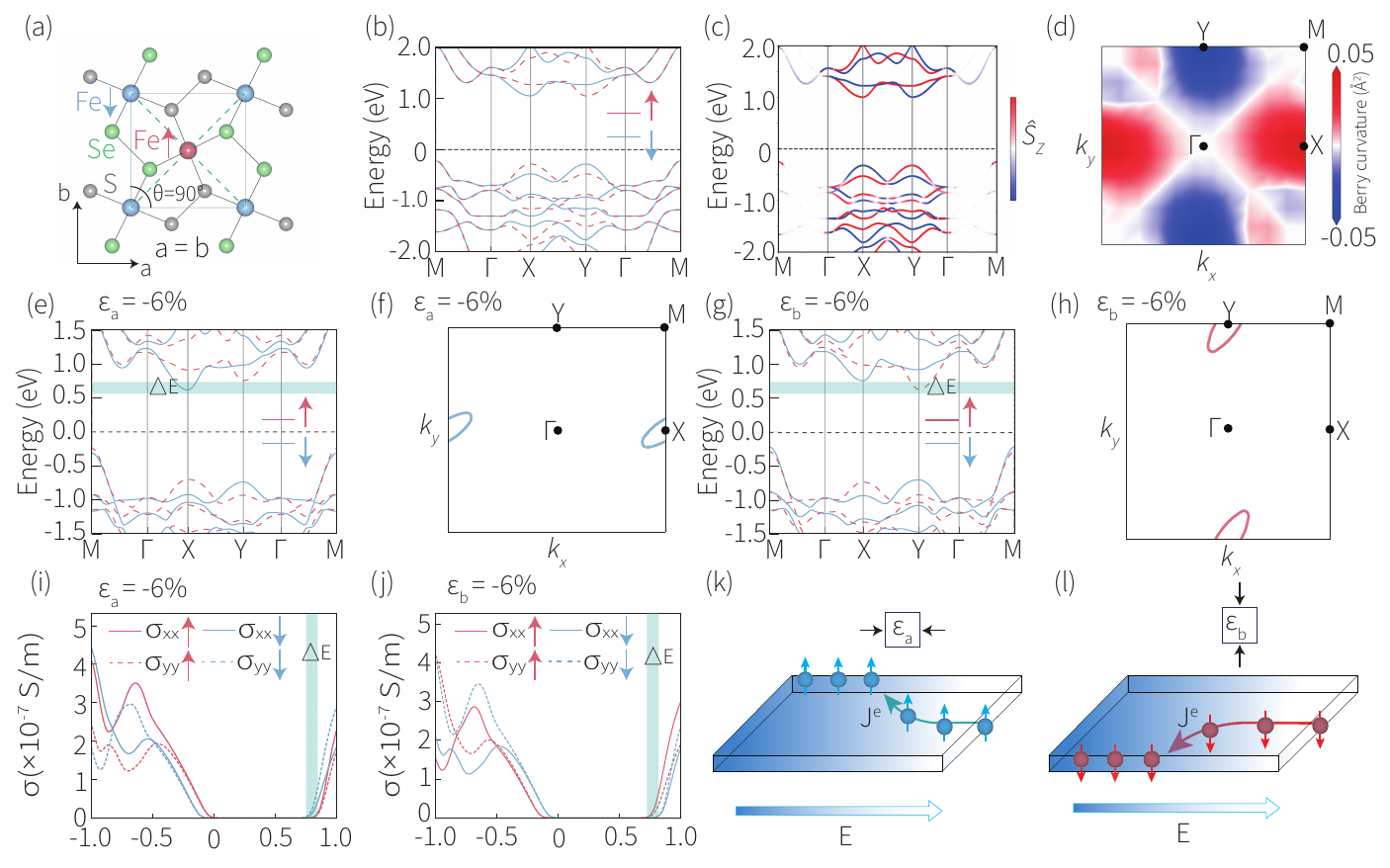}
  \caption{Schematic illustration of strain-tunable electronic structures, spin transport, and unconventional spin Hall effect in the FeSSe monolayer. (a) Top view of the FeSSe monolayer. (b) and (c) The band structures of FeSSe without and with SOC. (d) Distribution of Berry curvature summed for all valancebands for FeSSe. (e) and (f) The band structure of the FeSSe under 6$\%$ $x$-axis compressive strain and the corresponding Fermi Surface at the chemical potential of 0.75 eV, respectively. (g) and (h) The  band structure of the FeSSe under 6$\%$ $y$-axis compressive strain and the corresponding Fermi Surface at the chemical potential of 0.75 eV, respectively. (i) and (j) Transverse ($\sigma^{s}_{xx}$) and longitudinal ($\sigma^{s}_{yy}$) spin-resolved conductivity as a function of chemical potential at 300 K for the FeSSe under 6$\%$ $x$-axis compressive strain and under 6$\%$ y-axis compressive strain, respectively. (k) and (l) Present schematic diagrams of the unconventional spin Hall effect with an in-plane electric field under 6$\%$ compressive strain along the $x$-axis and $y$-axis for U$_{g}$ $>$ 0, respectively.}
  \label{Fig4}
\end{figure}

The pentagonal monolayer FeSSe exhibits a Type-III uniaxial-strain response. The crystal structure of FeSSe is presented in Figure~\ref{Fig4}a. FeSSe belongs to the orthorhombic space group \(Cmm2\) (No.~35) and spin Laue group \({}^2m{}^2m{}^1m\), with the full set of nontrivial symmetry operations connecting opposite-spin sublattices being:
$\mathcal{O}_{0}^{\uparrow\downarrow} = [M_{xy}, M_{\overline{xy}}].$
The spin-resolved band structure in Figure~\ref{Fig4}b shows nonrelativistic altermagnetic spin splitting, with low-energy valleys located near the X and Y points, thus exhibiting a spin-valley locking effect. The degeneracy of the bands along the $\Gamma$ $\rightarrow$ M path arises from the presence of the mirror planes \(M_{xy}\) and \(M_{\overline{xy}}\). In the unstrained structure, these two valleys are degenerate and both contribute to the band-edge electronic states.

The spin-orbit coupling (SOC)-included band structure of FeSSe is shown in Figure~\ref{Fig4}c, where the red and blue colors indicate the spin polarization \(\langle n\mathbf{k}|\hat{S}_z|n\mathbf{k}\rangle\) of the eigenstate \(|n\mathbf{k}\rangle\). The total Berry curvature of the FeSSe monolayer is shown in Figure~\ref{Fig4}d. Clear peaks with opposite signs appear at the X and Y valleys, revealing a valley-contrasting Berry curvature. The X- and Y-valley band edges carry opposite dominant spin characters, consistent with the spin-sublattice symmetry operations connecting the corresponding momentum-space states. Therefore, lifting the near-degeneracy of the X/Y valleys simultaneously selects conducting carriers with different spins.

Uniaxial compression introduces an anisotropic diagonal strain field, as shown in Eq \ref{eq:strain_tensors}. Unlike the shear response in \(\alpha\)-CoP\(_2\), this uniaxial perturbation removes both the symmetry relation that keeps the two spin-valley band edges equivalent and the nontrivial symmetry operations, \(M_{xy}\) and \(M_{\overline{xy}}\). As a result, no nontrivial opposite-spin sublattice-connecting operation remains to support the altermagnetic spin--momentum mapping. Consequently, the SML is broken, and the valley energies split:
$E_X(\varepsilon_\alpha) \neq E_Y(\varepsilon_\alpha), \alpha = x, y$.
Figures~\ref{Fig4}e,g show the band structures of the FeSSe monolayer under \(6\%\) compression along the \(x\)- and \(y\)-directions, respectively. When \(6\%\) compression is applied along the \(x\)-direction, the X-valley is lowered and becomes the conduction-band minimum. In contrast, compression along the \(y\)-direction places the conduction-band minimum at the Y-valley. As shown in Figures~\ref{Fig4}f,h, this feature is also captured by the Fermi surfaces obtained under strains applied along different directions. Because the selected valleys carry opposite spin characters, choosing different directions of the uniaxial strain reverses the spin polarity of the band-edge carriers.

The transport consequence is shown in Figure~\ref{Fig4}i,j. Within the selected energy window, only one spin channel contributes appreciably to the conductivity, while the opposite spin channel is suppressed. Since the conducting carriers are dominated by a single spin orientation, this gives rise to a nearly \(100\%\) spin-polarized current. Figure~\ref{Fig4}k,l schematically illustrate the unconventional spin Hall effect with an in-plane electric field after applying a gate voltage (\(U_g>0\)) under fixed strain conditions. Therefore, FeSSe exhibits a Type-III response, where uniaxial strain removes the valley-exchange constraint and converts the altermagnetic spin-valley locking into a mechanically switchable spin-polarized current.

Overall, these three compounds illustrate representative classes of SML evolution under different strain components. 
Furthermore, using the Ising model, we estimate the N\'eel temperatures of the three representative materials to be 320~K for \(\alpha\)-CoS\(_2\), 550~K for \(\alpha\)-CoP\(_2\), and 385~K for FeSSe, all above room temperature, as shown in Figure~S10 in the Supporting Information. These results indicate their potential for practical applications in room-temperature spintronic devices.

\section{Conclusion}
In summary, we have established a strain-resolved symmetry criterion for nonrelativistic altermagnetic spin splitting in pentagonal two-dimensional altermagnets. By tracking how continuous in-plane strain modifies the altermagnetic  SML, we classify 94 candidate systems from ten parent space groups into three response types: Type-I, where the SML is preserved; Type-II, where the SML is reconstructed; and Type-III, where the SML is destroyed by the removal of the relevant opposite-spin sublattice-connecting symmetry operations. Three representative compounds illustrate these symmetry-defined response classes. In \(\alpha\)-CoS\(_2\), ferroelastic switching represents a finite-strain Type-I domain response, in which the nontrivial symmetry operations connecting opposite-spin sublattices are preserved, while the spin-conductivity tensor is reversed by domain switching. In \(\alpha\)-CoP\(_2\), shear strain breaks a subset of the nontrivial opposite-spin sublattice-connecting operations, reconstructing the SML and producing a Type-II response that activates an off-diagonal spin-current response with \(\sigma_{xy}^{\uparrow} = -\sigma_{xy}^{\downarrow}\). In FeSSe, uniaxial strain removes all nontrivial operations connecting opposite-spin sublattices, leading to a Type-III response, the X/Y valley degeneracy is lifted, and a single-spin conducting channel is selected. These results not only provide guidance for the manipulation of two-dimensional orthorhombic altermagnets but also offer a pathway toward engineering their transport properties for future device applications.

\section{Computational Methods}
First-principles calculations were performed within the density functional theory (DFT) framework using the projected augmented wave (PAW) \cite{PAW} method implemented in the VASP package\cite{VASP}. The generalized gradient approximation (GGA) of the Perdew–Burke–Ernzerhof (PBE)\cite{PBE} was adopted to describe the exchange–correlation interactions between electrons. The kinetic cutoff energy was set to be 500 eV. The convergence threshold for self-consistent-field interaction was set to be 10$^{-6}$ eV. A $\Gamma$-centered 12$\times$12$\times$1 $k$-mesh grid was used for Brillouin zone (BZ) sampling. To avoid the interaction of periodic mirrors, the unit cell with a vertical vacuum space greater than 15 \AA \, was constructed. All geometric structures were fully optimized until the convergence tolerance, that is, the force on each atom was less than 0.01 eV/\AA. Correlation effects were considered within the GGA+U formalism introduced by Dudarev et al.\cite{DFTPU} using an effective on-site Hubbard parameter U$_{eff}$ 3.0, 3.5, 4.0, 4.0, and 3.3 eV for the 3$d$ states of V, Cr, Mn, Fe, and Co ions, respectively. The climbing image nudged elastic band (CINEB) method\cite{CINEB} was used to determine the energy barriers of the various kinetic processes. The phonon spectra were calculated by density functional perturbation theory\cite{DFPT1, DFPT2}, and the data were derived by phonopy\cite{PHONOPY}. VASPKIT\cite{VASPKIT} was adopted to finish the pre-processing and post-processing of electronic structure calculations. The Berry curvature is defined as:
\[
\Omega_{n k} = -2 \operatorname{Im} \sum_{n^{\prime} \neq n} \frac{\langle n \mathbf{k} | v_x | n^{\prime} \mathbf{k} \rangle \langle n^{\prime} \mathbf{k} | v_y | n \mathbf{k} \rangle}{(\omega_{n^{\prime}} - \omega_n)^2},
\tag{4}
\]
where $v_{x}$ and $v_{y}$  are the velocity operators, and E$_{n}$= $\hbar$$\omega$$_{n}$ is the energy of the state $\left\langle\psi_{nk}\right|$  . The total Berry curvature, obtained by summing over all occupied valence bands, is given by $\Omega$(k)= ${\textstyle \sum_{n\in occ }}$ .Based on the VASPBERRY code, we computed the Berry curvature\cite{kim2022berry}.
According to the Boltzmann transport equation, under the relaxation time  $\tau$ approximation, the conductivity can be written as \cite{Boltz}:
\[
\sigma_{ij}^s(\epsilon_F) = -\frac{e^2 \tau}{8\pi^3 \hbar^2} \sum_n \int \frac{\partial \varepsilon_{n\mathbf{k}}^s}{\partial k_i} \frac{\partial \varepsilon_{n\mathbf{k}}^s}{\partial k_j} \frac{\partial f^0}{\partial \varepsilon_{n\mathbf{k}}^s} \, d^3k,
\tag{5}
\]

where $f^0$($\varepsilon_{nk}^s$, $\epsilon_F$)  is the Fermi distribution function, expressed in terms of the eigenvalue $\varepsilon_{nk}^s$ of the 
$n$-th band and the Fermi energy. $\hbar$ is the reduced Planck constant; $k_{ij}$ is the wave vector in Cartesian coordinates. Following the implementation in the WANNIER90 code \cite{Wannier90} based on the maximally localized Wannier function basis set, we performed the conductivity calculations. The temperature was set to 300 K in the Fermi distribution function, and a k-point mesh of 1000 $\times$ 1000 $\times$1 was employed. The anisotropic spin polarization $\text{SP}_{\hat{n}}$ is defined as \cite{Ansc}:
\[\text{SP}_{\hat{n}} = \frac{\sigma_{\hat{n}\hat{n}}^{\uparrow} - \sigma_{\hat{n}\hat{n}}^{\downarrow}}{\sigma_{\hat{n}\hat{n}}^{\uparrow} + \sigma_{\hat{n}\hat{n}}^{\downarrow}}, \tag{6}\]
Consider the spin-polarized conductivity $\sigma_{\hat{n}\hat{n}}^{s}$ along the direction $\hat{n} = (\sin\theta \cos\phi, \sin\theta \sin\phi, \cos\theta)$ (where $\theta$ is the polar angle and $\phi$ is the azimuthal angle). Since the system is a two-dimensional material confined in the $xy$-plane, the $z$-direction components vanish. With the aid of directional derivatives, one finds:
\[
\sigma_{\hat{n}\hat{n}}^{s} = \sigma_{xx}^{s} \sin^2\theta \cos^2\phi + \sigma_{yy}^{s} \sin^2\theta \sin^2\phi + \sigma_{xy}^{s} \sin^2\theta \sin(2\phi),
\tag{7}
\]
where $\theta = \pi/2$ (i.e., $\sin\theta = 1$) for in-plane transport, further simplifying the expression.
Here, $\sigma_{\hat{n}\hat{n}}^{s}$ denotes the conductivity along the $\hat{n}$ direction for spin channel $s$ ($s = \uparrow, \downarrow$), and $\sigma_{ij}^{s}$ are the components of the conductivity tensor for spin channel $s$.
Monte Carlo simulations were carried out using the Heisenberg Hamiltonian in the Ising model, aiming to estimate the N\'{e}el temperature $T_{\mathrm{N}}$. The simulations were performed on a $32 \times 32 \times 1$ supercell using the MCSLOVER package \cite{mcsol}, with $4 \times 10^{3}$ Monte Carlo steps for thermal equilibration at each temperature.









\section{Authors contributions}
$\dagger$ These authors contributed equally to this work.
\section*{Acknowledgements}
\textbf{Acknowledgements} \par 
This work was financially supported by the National Natural Science Foundation of China (Grant No. 12304165, 12304086, 12564017), the Natural Science Foundation of Inner Mongolia Autonomous Region (Grant No. 2026QB016), and the Startup Project of Inner Mongolia University (Grant No. 21200-5223733).

\section*{Supporting information}

\begin{itemize}
  \item Figure S1, Table S1, Note S1 and Note S2: the 10 altermagnetism-capable space groups and their symmetry analyses
  \item Figure S2 and Figure S3: the phonon spectra and band structures
  \item Figure S4 Table S2 and Table S3: ground state configurations and magnetic energy comparisons
  \item Figure S5: the band structures and corresponding transition states of other representative ferroelastic materials 
  \item Figures S6–S9: strain-controlled valley polarization and SOC band structures for representative pentagonal materials
  \item Figure S10: Simulated N\'{e}el temperature $T_{\mathrm{N}}$
  \item Figure S11: comparison of band structures between DFT and Wannier90-based calculations
\end{itemize}

\bibliography{MSP-afm}

\newpage



  
  

\end{document}